\documentclass[11pt]{article}
\usepackage{graphicx}
\usepackage{amsmath,amssymb}
\def\1{{\bf 1}}

\def\be{\begin{equation}}
\def\ee{\end{equation}}

\begin{document}
\noindent {\small Invited Book Chapter for ``Neurodegenerative Diseases / Book 2, Raymond Chuen-Chung Chang (eds.), INTECH Open Access Publisher, 2011, ISBN 979-953-307-672-9, 1st Version".}\\ 
{\bf--------------------------------------------------------------------------------------------------- }\\

\bigskip

\noindent {\large \bf Computational Studies of the Structural Stability of Rabbit Prion Protein Compared to Human and Mouse Prion Proteins}\\

\bigskip

\noindent {\Large Jiapu Zhang}\\

\noindent Centre in Informatics and Applied Optimization\&\\ 
Graduate School of Sciences, Informatics Technology and Engineering,\\  
University of Ballarat, Mount Helen, VIC 3353, Australia.\\
Emails: j.zhang@ballarat.edu.au, jiapu\_zhang@hotmail.com,\\ 
Telephones: 61-4 2348 7360 (mobile), 61-3-5327 9809 (office), Fax: 61-3-5327 9289\\

\noindent {\bf Abstract}\\ 
Prion diseases are invariably fatal and highly infectious neurodegenerative diseases affecting humans and animals. The neurodegenerative diseases such as Creutzfeldt-Jakob disease, variant Creutzfeldt-Jakob diseases, Gerstmann-Str$\ddot{a}$ussler-Scheinker syndrome, Fatal Familial Insomnia, Kuru in humans, scrapie in sheep, bovine spongiform encephalopathy (or 'mad-cow' disease) and chronic wasting disease in cattle belong to prion diseases. By now there have not been some effective therapeutic approaches to treat all these prion diseases. In 2008, canine mammals including dogs (canis familials) were the first time academically reported to be resistant to prion diseases (Vaccine 26: 2601-2614 (2008)). Rabbits are the mammalian species known to be resistant to infection from prion diseases from other species (Journal of Virology 77: 2003-2009 (2003)). Horses were reported to be resistant to prion diseases too (Proceedings of the National Academy of Sciences USA 107: 19808-19813 (2010)). By now all the NMR structures of dog, rabbit and horse prion proteins had been released into protein data bank respectively in 2005, 2007 and 2010 (Proceedings of the National Academy of Sciences USA 102: 640-645 (2005), Journal of Biomolecular NMR 38:181 (2007), Journal of Molecular Biology 400: 121-–128 (2010)). Thus, at this moment it is very worth studying the NMR molecular structures of horse, dog and rabbit prion proteins to obtain insights into their immunity prion diseases. 

The author found that dog and horse prion proteins have stable molecular dynamical structures whether under neutral or low pH environments, but rabbit prion protein has stable molecular dynamical structures only under neutral pH environment. Under low pH environment, the stable $\alpha$-helical molecular structures of rabbit prion protein collapse into $\beta$-sheet structures. This article focuses the studies on rabbit prion protein (within its C-terminal NMR, Homology and X-ray molecular structured region RaPrP$^\text{C}$ (120-230)), compared with human and mouse prion proteins (HuPrP$^\text{C}$ (125-228) and MoPrP$^\text{C}$ (124-226) respectively). The author finds that some salt bridges contribute to the structural stability of rabbit prion protein under neutral pH environment.\\


\section{Introduction}
As we all know, the disease infectious prions PrP$^\text{{Sc}}$ are rich in $\beta$-sheets (about 43\% $\beta$-sheet) \cite{griffith1967} and the normal cellular prions PrP$^\text{{C}}$ are predominant in $\alpha$-helices (42\% $\alpha$-helix, 3\% $\beta$-sheet) \cite{pan_etal1993}. Prion diseases are believed caused by the conversion of normal PrP$^\text{{C}}$ to abnormally folded PrP$^\text{{Sc}}$, and prion diseases are so-called protein `structural conformational' diseases. Thus, we may study the molecular structures of prion proteins to obtain some insights of prion diseases. Human prion diseases include the Creutzfeldt-Jakob disease, variant Creutzfeldt-Jakob diseases, Gerstmann-Str$\ddot{a}$ussler-Scheinker syndrome, Fatal Familial Insomnia, and Kuru. The NMR solution structure of the human prion protein (1QLX.pdb) was released into the PDB bank (www.pdb.org) in 1999 and last modified in 2009 \cite{zahn_etal2000}. Mice are popular experimental laboratory animals and the NMR structure of mouse prion protein (1AG2.pdb) was released into the PDB bank in 1997 and last modified in 2009 \cite{riek_etal1996}. Rabbits, dogs and horses were reported to be resistant to prion diseases \cite{vorberg_etal2003,polymenidoua_etal2008,perez_etal2010} and by the end of 2010 their NMR structures (2FJ3.pdb, 1XYK.pdb, and 2KU4.pdb respectively) were completed to release into the PDB bank \cite{wen_etal2010a,lysek_etal2005,perez_etal2010}. The X-ray structure of rabbit prion protein (3O79.pdb) was released into PDB bank in 2010 too \cite{khan_etal2010}. At this moment it is very worth studying these molecular structures of horse, dog, rabbit, human and mouse prion proteins to reveal some secrets of prion diseases. The author found that dog and horse prion proteins have stable molecular dynamical structures whether under neutral or low pH environments \cite{zhangl2011, zhang2011a}, but rabbit prion protein has stable molecular dynamical structures only under neutral pH environment \cite{zhangl2011, zhang2011a, zhang2010, zhang2011b}. Under low pH environment (at 450 K and 350 K), among all the prion proteins above-mentioned only for the rabbit prion protein, its stable $\alpha$-helical molecular structures collapse into $\beta$-sheet structures \cite{zhangl2011, zhang2011a, zhang2010, zhang2011b}. The conversion of disease PrP$^\text{{Sc}}$ from normal PrP$^\text{{C}}$ is just involving `conformational change' from predominantly $\alpha$-helical protein to one rich in $\beta$-sheet structure. This article specially focuses on the rabbit prion protein to obtain some insights into the immunity of rabbits to prion diseases.\\

For the rabbit prion protein, we have its NMR and X-ray structures (2FJ3.pdb and 3O79.pdb respectively). Early in 2004, Epa \cite{zhangev2006} made molecular modeling of a homology structure (denoted 6EPA.pdb) for rabbit prion protein RaPrP$^{\text{C}}$(120-229), which was constructed using the NMR structure of the human prion protein (1QLX.pdb) as the template. Besides all these structures of rabbit prion protein, the knowledge on their conformational evolution/dynamics is considered essential to understand rabbit prion and the molecular modeling (MM) and molecular dynamics (MD) approach takes advantage beyond the experimental limit. In Section 2, this article first briefly reviews the main MD results of the homology MM structure \cite{zhangev2006} at 500 K and of the wild-type NMR structure at 450 K \cite{zhang2011b}, compared with the MD of the wild-type NMR structures of human (HuPrP$^\text{C}$ (125-228), 1QLX.pdb) and mouse (MoPrP$^\text{C}$ (124-226), 1AG2.pdb) prion proteins. Section 3 will present the wild-type NMR rabbit, human and mouse MD comparisons at 350 K. Because the X-ray structure 3O79.pdb was produced differently from the NMR structure 2FJ3.pdb and the Homology structure 6EPA.pdb, we will not do their MD comparisons; however, in Section 4, we will give detailed sequence and structure alignment analysis of all these three rabbit prion structures compared with human and mouse wild-type structures. Section 5 gives some concluding remarks on rabbit prion protein and prion diseases.

\section{MD Reviews On the 500 K Homology Rabbit Prion Protein and the 450 K NMR Rabbit Prion Protein}
\subsection{The Homology Rabbit Prion Protein at 500 K \cite{zhangev2006}}
\begin{figure}[h!]
\centerline{
\includegraphics[width=5.6in]{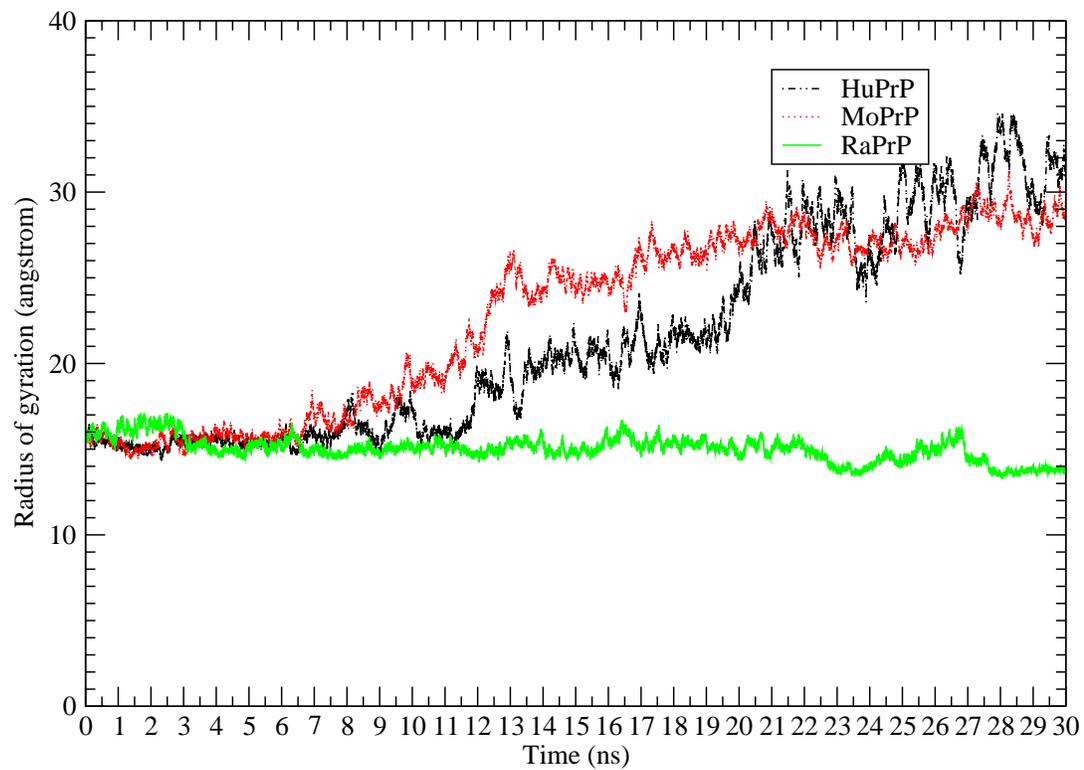}
} 
\caption{RMSD and Radius Of Gyration graphs at 500 K neutral pH value.}
\label{homology_RMSD_Radius}
\end{figure}

Zhang et al. \cite{zhangev2006} studied the MD of RaPrP$^{\text{C}}$(120-229) homology structure (6EPA.pdb). The MD simulations used Amber 8 \cite{amber8} PMEMD program, with explicit water at different temperatures and pH values. The simulation conditions are listed in Table \ref{csiro}.
\begin{table}[h!]
\begin{center}                 
\caption{Simulation conditions for the homology model.}
\label{csiro} {\tiny
\begin{tabular}{cccccc}
\\ \hline      
pH value   &Specie  &Truncated octahedral box (angstroms) &Total atoms &Ions added &TIP3P Waters added\\ \hline
Neutral pH &HuPrP   &75.868                               &19484       &3Na+       &5929\\ \cline{2-6}
           &MoPrP   &67.447                               &13422       &2Na+       &3918\\ \cline{2-6}
           &RaPrP   &78.130                               &21469       &2Na+       &6572\\ \hline
Low pH     &HuPrP   &74.834                               &18530       &16Cl-      &5599\\ \cline{2-6}
           &MoPrP   &67.335                               &13208       &14Cl-      &3836\\ \cline{2-6}
           &RaPrP   &80.896                               &23847       &14Cl-      &7354\\ \hline    
\end{tabular}}
\end{center}
\end{table}
The RMSD (root mean square deviation) and radius of gyration results are shown in Fig. \ref{homology_RMSD_Radius}.\\

\begin{figure}[h!]
\centerline{
\includegraphics[width=6.5in]{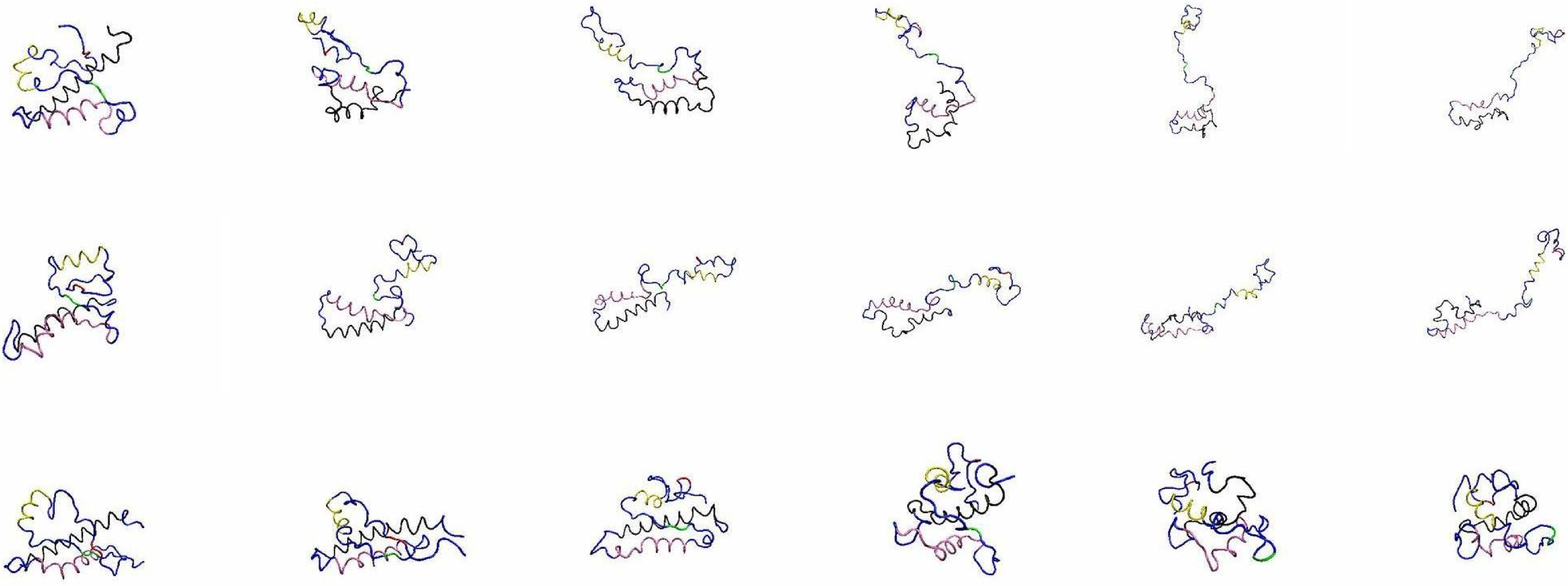}
} 
\caption{Snapshots of HuPrP, MoPrP and RaPrP (from up to down) at 500 K neutral pH value, at 5ns, 10ns, 15ns, 20ns, 25ns, and 30ns respectively (from left to right).}
\label{homology_snapshots}
\end{figure}

\centerline{$<\text{Fig. \ref{homology_RMSD_Radius}}>$}
\vskip 0.5cm
\noindent We may see that in Fig. \ref{homology_RMSD_Radius} rabbit prion protein has more stable structural dynamical behavior compared to the human and mouse prion proteins at 500 K under neutral pH environment. This is also shown in Fig. \ref{homology_snapshots} of snapshots for human, mouse and rabbit prion proteins at 5ns, 10ns, 15ns, 20ns, 25ns, and 30ns respectively.\\

\begin{figure}[h!]
\centerline{
\includegraphics[width=5.6in]{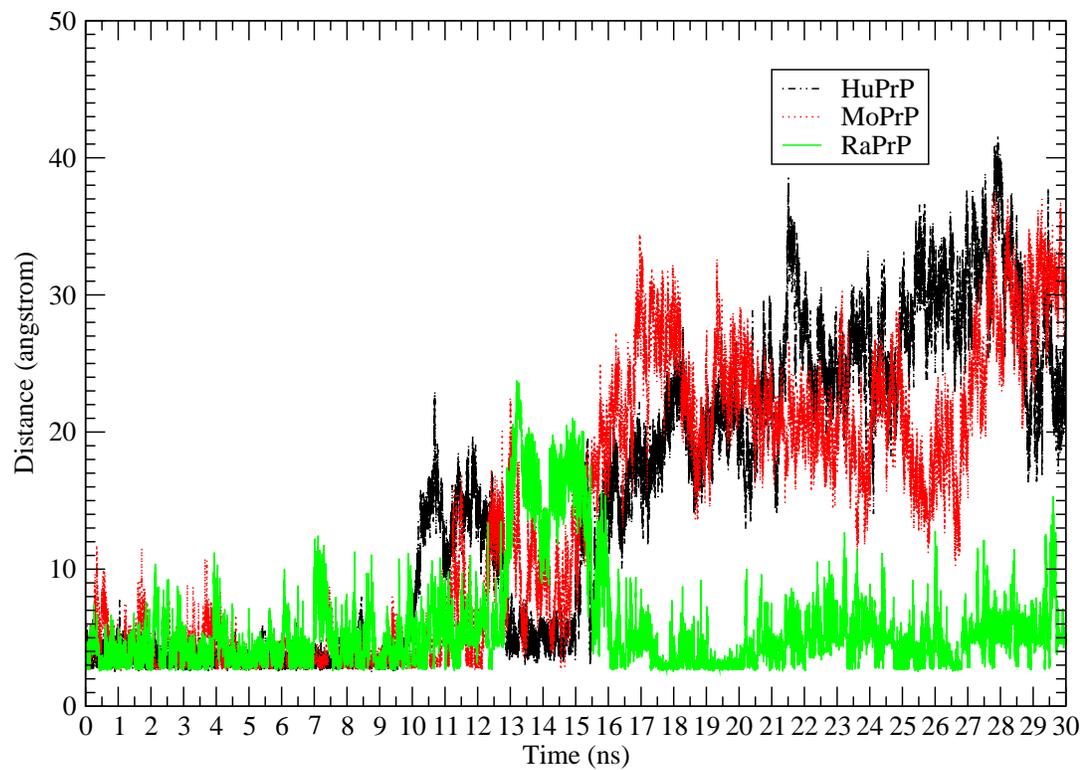}
} 
\caption{Distance between D177.OD1 and R163.NE for RaPrP, between D178.OD1 and R164.NE for HuPrP and MoPrP, at 500 K neutral pH value.}
\label{homology_saltbridge}
\end{figure}

\centerline{$<\text{Fig. \ref{homology_snapshots}}>$}
\vskip 0.5cm
\noindent Fig. \ref{homology_snapshots} shows that the helices of HuPrP and MoPrP were unfolded but RaPrP still keeps the helical structures at 500 K under neutral pH environment. Under low pH environment at 500 K, these helical structures of RaPrP were unfolded. One of the reasons of the rabbit prion protein unfolding is due to the remove of the salt bridges such as N177-R163 (Fig. \ref{homology_saltbridge}).\\

\centerline{$<\text{Fig. \ref{homology_saltbridge}}>$}
\vskip 0.5cm 
\noindent We may see in Fig. \ref{homology_saltbridge} the salt bridge / hydrogen bond between Arginine 163 and Aspartic acid 177 is conserved through a large part of the simulations and contributes to the protein stability of rabbit prion protein structure. Simulations at low pH value, where this salt bridge is absent, show RMSD and radius of gyration values for the rabbit prion protein to be of the same magnitude as the human and mouse prion proteins.

\subsection{The NMR Rabbit Prion Protein at 450 K \cite{zhang2011b}}
\begin{figure}[h!]
\centerline{
\includegraphics[width=4.2in,angle=90]{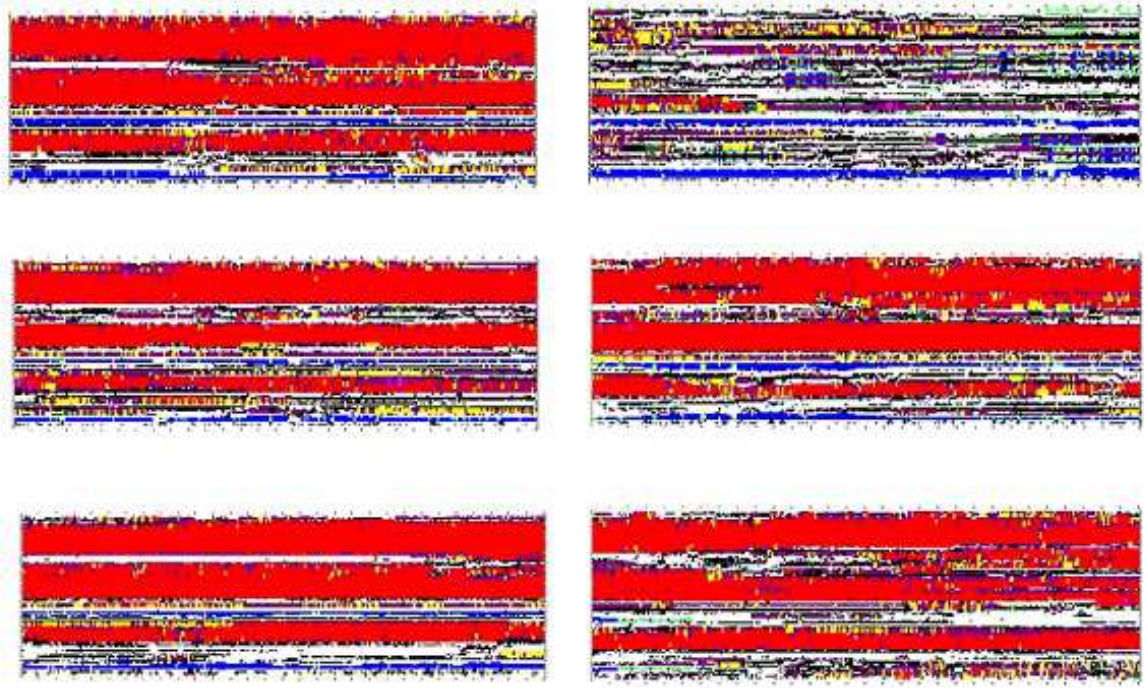}
} 
\caption{Secondary structures of rabbit, human and mouse prion proteins (from up to down) at 450 K under neutral and low (from left to right) pH environments (red: $\alpha$-helix, pink: $\pi$-helix, yellow: $3_{10}$-helix, green: $\beta$-bridge, blue: $\beta$-sheet, purple: Turn, Black: Bend; x-axis: time (0-30 ns), y-axis: residue numbers).}
\label{NMR_secondarystructure}
\end{figure}

Zhang \cite{zhang2011b} did the MD studies on the NMR rabbit prion protein RaPrP$^{\text{C}}$(124-228) (2FJ3.pdb) at 450 K under both neutral and low pH environments for the simulations of 20 ns. Zhang \cite{zhang2011b} found that ``the secondary structures under low pH environment at 450 K have great differences between rabbit prion protein and human and mouse prion proteins: the $\alpha$-helices of rabbit prion protein were completely unfolded and began to turn into $\beta$-sheets but those of human and mouse prion proteins were not changed very much. These results indicate the C-terminal region of RaPrP$^{\text{C}}$ has lower thermostability than that of HuPrP$^{\text{C}}$ and MoPrP$^{\text{C}}$. Under the low pH environment, the salt bridges such as D177-R163, D201-R155 were removed (thus the free energies of the salt bridges changed the thermostability) so that the structure nearby the central helices 1–3 was changed for rabbit prion protein" \cite{zhang2011b}. The author continued his MD simulations for another 10 ns. The secondary structures for the MD simulations of 30 ns (Fig. \ref{NMR_secondarystructure}) shows the same conclusion as that of \cite{zhang2011b}.\\

\centerline{$<\text{Fig. \ref{NMR_secondarystructure}}>$}
\vskip 0.5cm 
\noindent We may say that the salt bridges such as D177-R163, N201-R155 contribute to the structural stability of wild-type rabbit prion protein (Fig. \ref{NMR_secondarystructure}). At 450 K, whether in neutral or in low pH environments, the $\alpha$-helical secondary structures of dog prion protein have not changed for the long 30 ns' simulations \cite{zhangl2011}; at 350 K, horse prion protein has the same molecular structural dynamics \cite{zhang2011a} during the 30 ns' long simulations.

\section{350 K} 
350 K might be a practical temperature for some experimental laboratory works. Zhang \cite{zhang2011c} did MD simulations for wild-type rabbit, dog and horse prion NMR structures at 350 K. The findings of 350 K are: ``dog and horse prion proteins have stable molecular structures whether under neutral or low pH environments. Rabbit prion protein has been found having stable molecular structures under neutral pH environment, but without structural stability under low pH environment. Under low pH environment, the salt bridges such as D177-R163 were broken and caused the collapse of the stable $\alpha$-helical molecular structures". Here the MD simulations are done for wild-type human and mouse prion proteins in the use of the same Materials and Methods as in \cite{zhang2011c}.\\

\begin{figure}[h!]
\centerline{
\includegraphics[width=2.8in]{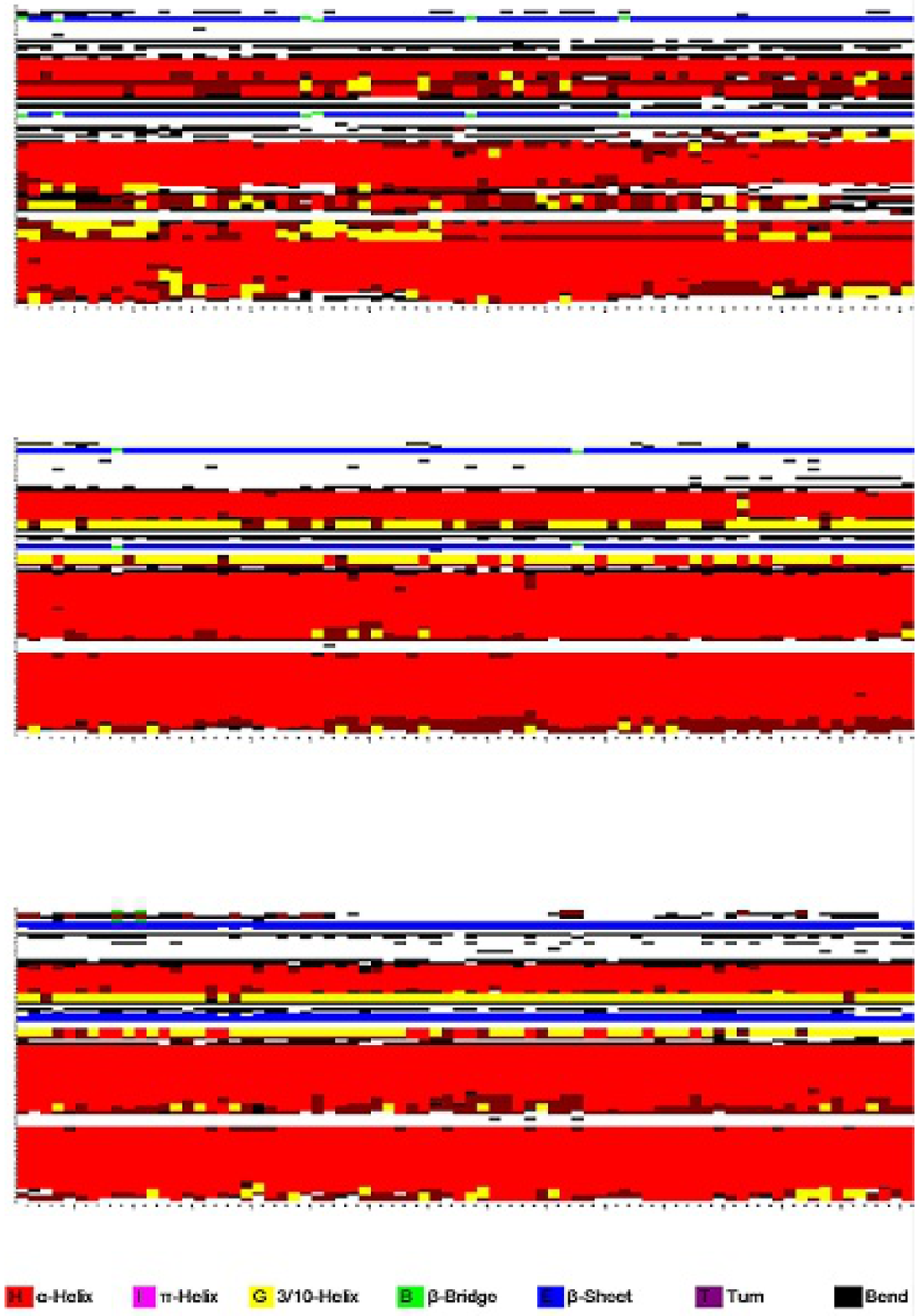} \quad
\includegraphics[width=2.8in]{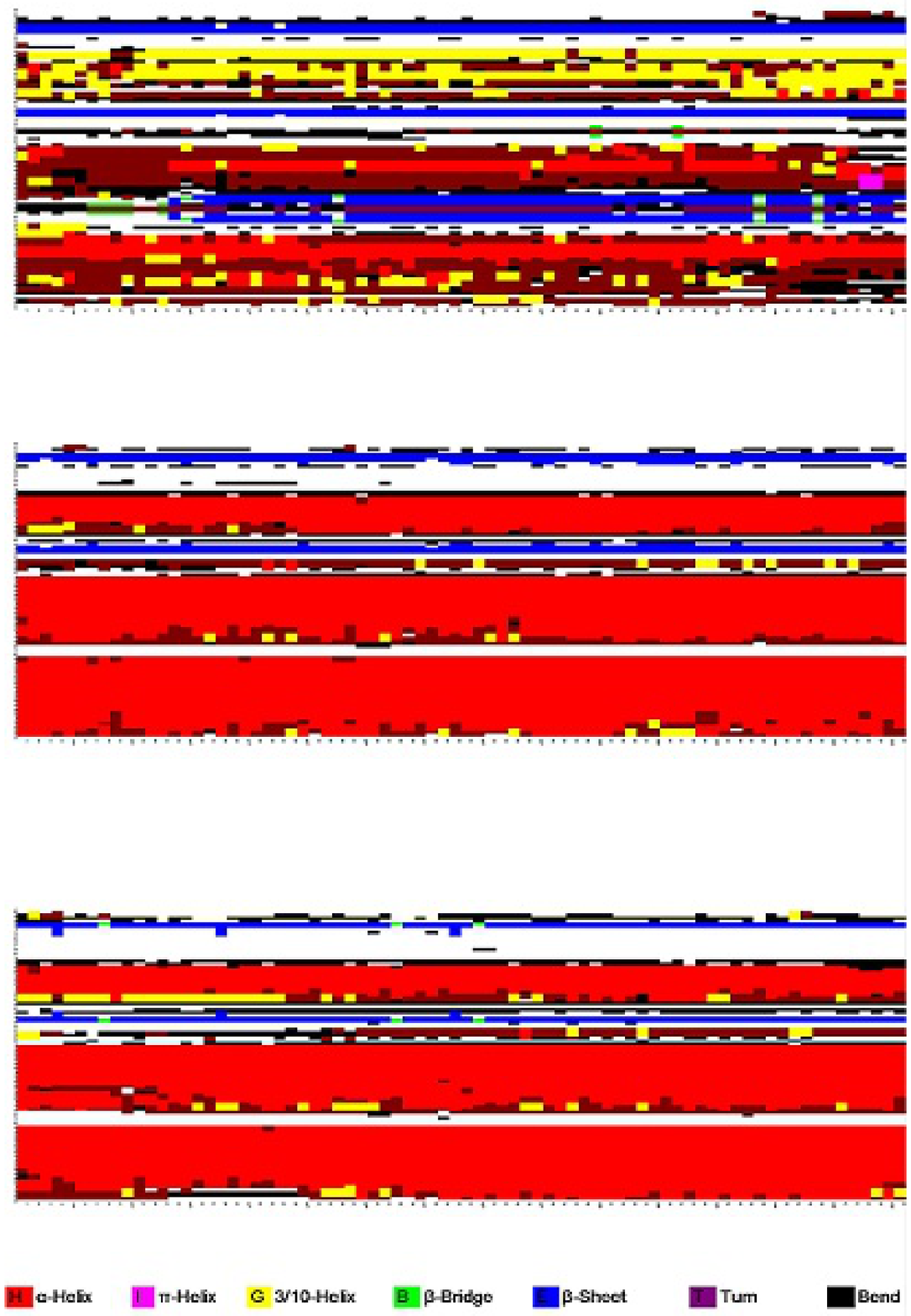}
}
\caption{Secondary structures of rabbit, human and mouse prion proteins (from up to down) at 350 K under neutral to low pH
environments (from left to right) (X-axis: 0 ns - 30 ns (from left to right), Y-axis: residue numbers 124 - 228 / 125 - 228 / 124 - 226 (from up to down)).}
\label{NMR_350K_secondary_structures}
\end{figure}

\centerline{$<\text{Fig. \ref{NMR_350K_secondary_structures}}>$}
\vskip 0.5cm 
\noindent Seeing Fig. \ref{NMR_350K_secondary_structures}, we know that at 350 K human and mouse prion proteins have stable molecular structures whether under neutral or low pH environments.\\

\begin{figure}[h!]
\centerline{
\includegraphics[width=6.5in]{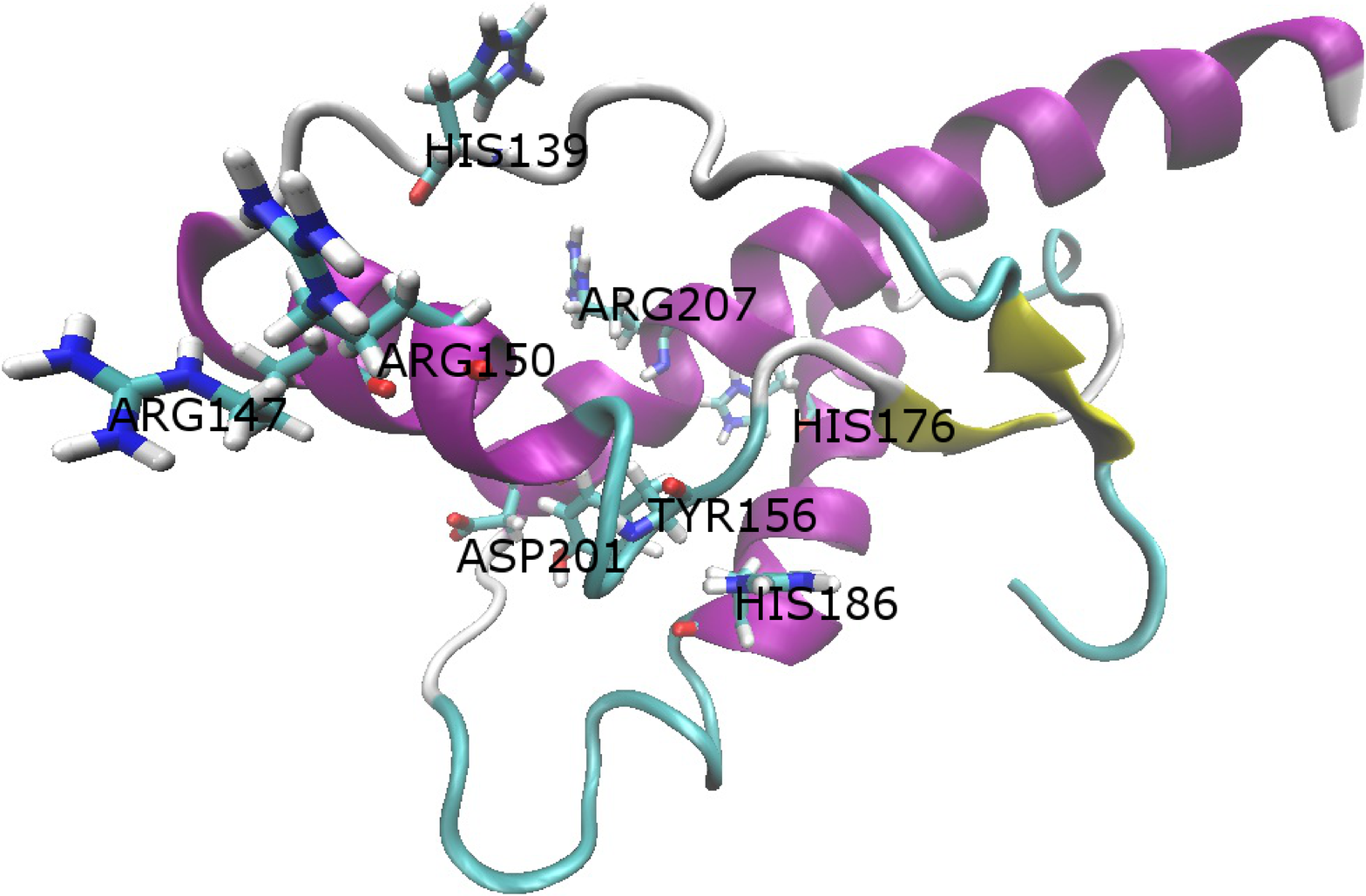}
} 
\caption{Some special salt bridges ARG207-HIS176, TYR156-HIS186, HIS139-ARG150, ASP201-ARG147, ASP201-ARG150, ASP201-HIS186, ARG155-GLU151 of wild-type NMR rabbit prion protein at 350 K.}
\label{rabbit_saltbridges_350K}
\end{figure}

Clearly the following salt bridges play an important role to the NMR structural stability of rabbit prion protein: (1) GLU210-ARG207-GLU206-LYS203 (99.78\%, 88.85\%, 82.74\%, H3-H3), GLU210-HIS176 (74.31\%, H3-H2), GLU206-HIS176 (57.10\%, H3-H2), ARG207-HIS176 (0.52\%, H3-H2), ASP177-ARG163 (19.54\%, H2-S2); (2) ARG150-ASP146-ARG147-ASP143 (91.38\%, 100\%, 86.43\%, H1-H1), HIS139-ARG150 (50.96\%), HIS139-ASP146 (92.62\%); (3) ASP201-ARG155 (10.07\%, H3-H1), ASP201-ARG150 (2.61\%, H3-H1), ASP201-ARG147 (0.01\%, H3-H1), ASP201-HIS186 (0.50\%, H3-H2); and(4) ARG155-ASP201 (10.07\%, H1-H3), TYR156-HIS186 (H1-H2, 71.69\%), ARG155-GLU151 (20.70\%, H1-H1), ARG155-GLU195 (0.06\%), where H1, H2, H3 denote the $\alpha$-helix 1, 2, 3 respectively, S1, S2 denote the $\beta$-strand 1 and 2 respectively, and `\%' denotes the percentage during the whole simulation of 30 ns. Compared with human, mouse, dog and horse NMR prion proteins, rabbit NMR prion protein has some special salt bridges which contribute to its structural stability at 350 K during the simulation of 30 ns (Fig. \ref{rabbit_saltbridges_350K}) (human, mouse, dog and horse NMR prion proteins have not these salt bridges).\\

\centerline{$<\text{Fig. \ref{rabbit_saltbridges_350K}}>$}
\vskip 0.5cm

\section{Alignment Analyses}
\begin{figure}[h!]
\centerline{
\includegraphics[width=6.5in]{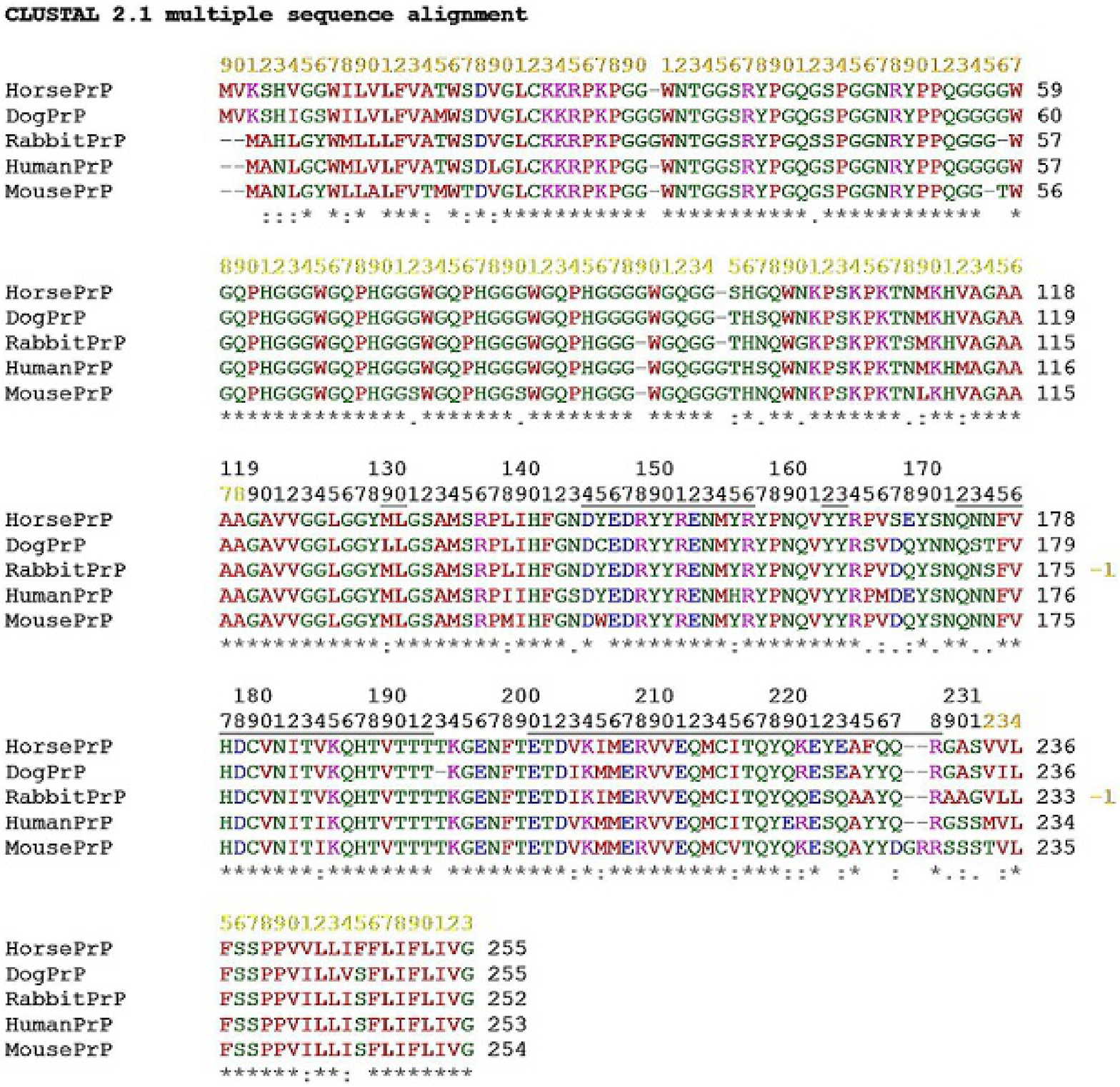}
} 
\caption{Horse, dog, rabbit, human and mouse prion protein sequence alignment.}
\label{prion_alignments}
\end{figure}

We make the sequence alignment of PrP from horse, dog, rabbit, human and mouse protein (Fig. \ref{prion_alignments}).\\

\begin{figure}[h!]
\centerline{
\includegraphics[width=2.0in]{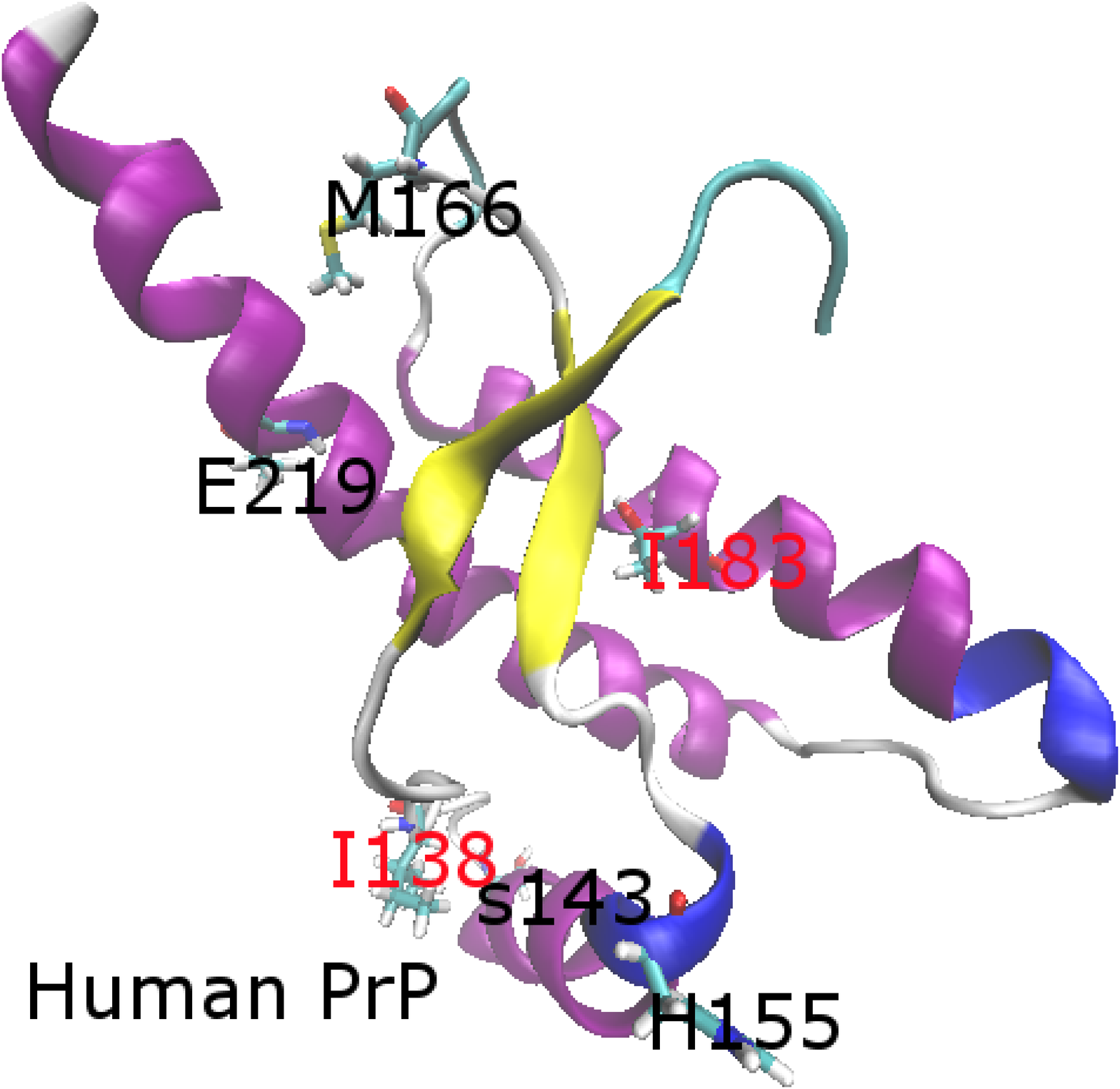}
\includegraphics[width=2.0in]{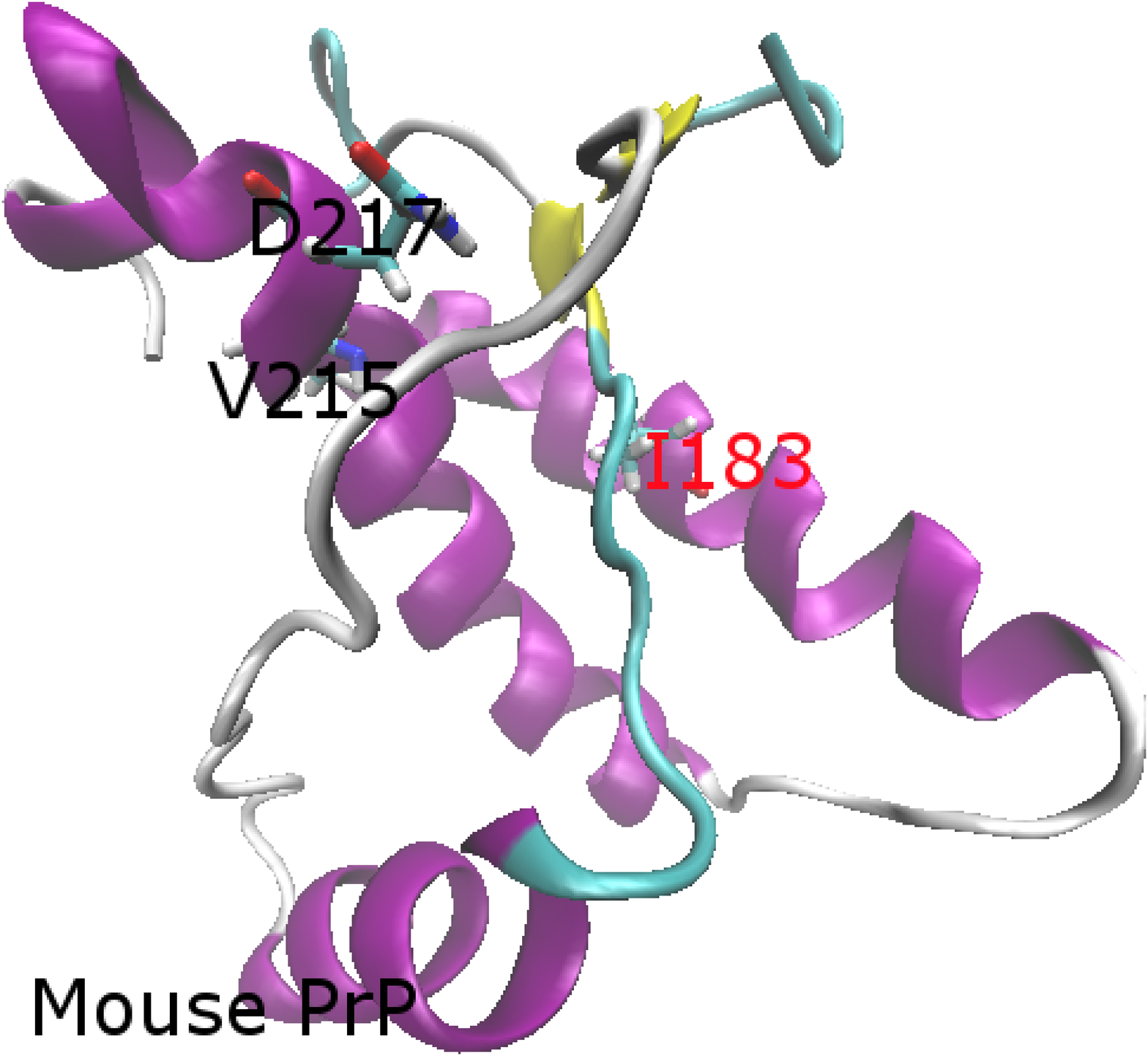}
}
\caption{Special residues owned only by HoPrP, DoPrP, RaPrP, HuPrP, and MoPrP respectively.}
\label{aa_special prions}
\end{figure}

\centerline{$<\text{Fig. \ref{prion_alignments}}>$}
\vskip 0.5cm 
\noindent In Fig. \ref{prion_alignments}, ``*" means that the residues in that column are identical in all sequences in the alignment, ``:" means that conserved substitutions have been observed, ``." means that semi-conserved substitutions are observed, the RED color takes place at small (small+ hydrophobic (incl.aromatic-Y)) residues, the BLUE color takes place at acidic residues, the MAGENTA color takes place at Basic-H residues,GREEN color takes place at Hydroxyl+sulfhydryl+amine+G residues, and Grey color takes place at unusual amino/imino acids etc.. For the structural domain, in Fig. \ref{prion_alignments} we can see some special residues listed in Table \ref{alignment_aa} for horse, dog, human and mouse prion proteins, which might contribute to characters of each structure respectively.  
\begin{table}[h!]
\begin{center}                 
\caption{Alignment analysis of special residues for HoPrP, DoPrP, HuPrP, and MoPrP.}
\label{alignment_aa} {\tiny
\begin{tabular}{ll} 
 \hline      
Horse       &S167 (others are D), Y222 (others are S), Q226 (others are Y), V241 (others are I), F245 (others are S)\\ \hline
Dog         &L129 (others are M), S165 (others are P), N170 (others are S), S173 (others are N), V244 (others are I)\\ \hline
Human       &I138 (immunities are L), S143 (others are N), H155 (others are Y), M166 (others are V), I183 (immunities are V),\\
            &E219 (others are Q), S230 (immunities are A)\\ \hline
Mouse       &I183 (immunities are V), V215 (others are I), D217 (others are Q), S230 (immunities are A)\\ \hline 
\end{tabular}
}
\end{center}
\end{table}
\noindent Rabbits differ from horses, dogs, humans and mice at: S173 (N174 for horse, T174 for dogs, N174 for humans and mice), Q219 (K220 for horses and humans, R220 for dogs and mice), A224 (F225 for horses, Y225 for dogs, humans and mice), L232 (I233 for dogs, V233 for horse, humans and mice), and G228 (others are S229). For rabbits, at positions 89 and 97  the residues are special from all others (G89 (others are N90), S97 (others are N98)).  These special residues are illuminated in Fig. \ref{aa_special prions}. Some recent researches are focusing on the loop between $\beta$2 and $\alpha$2, i.e. PrP(164-171) \cite{apostol_etal2011, fernandez-funez_etal2011, khan_etal2010, perez_etal2010, wen_etal2010a, wen_etal2010b}; we may see in Fig. \ref{aa_special prions} that the immune animals horses, dogs and rabbits have some residues in this loop different from humans and mice.\\ 

\centerline{$<\text{Fig. \ref{aa_special prions}}>$}
\vskip 0.5cm 

\begin{figure}[h!]
\centerline{
\includegraphics[width=3.5in]{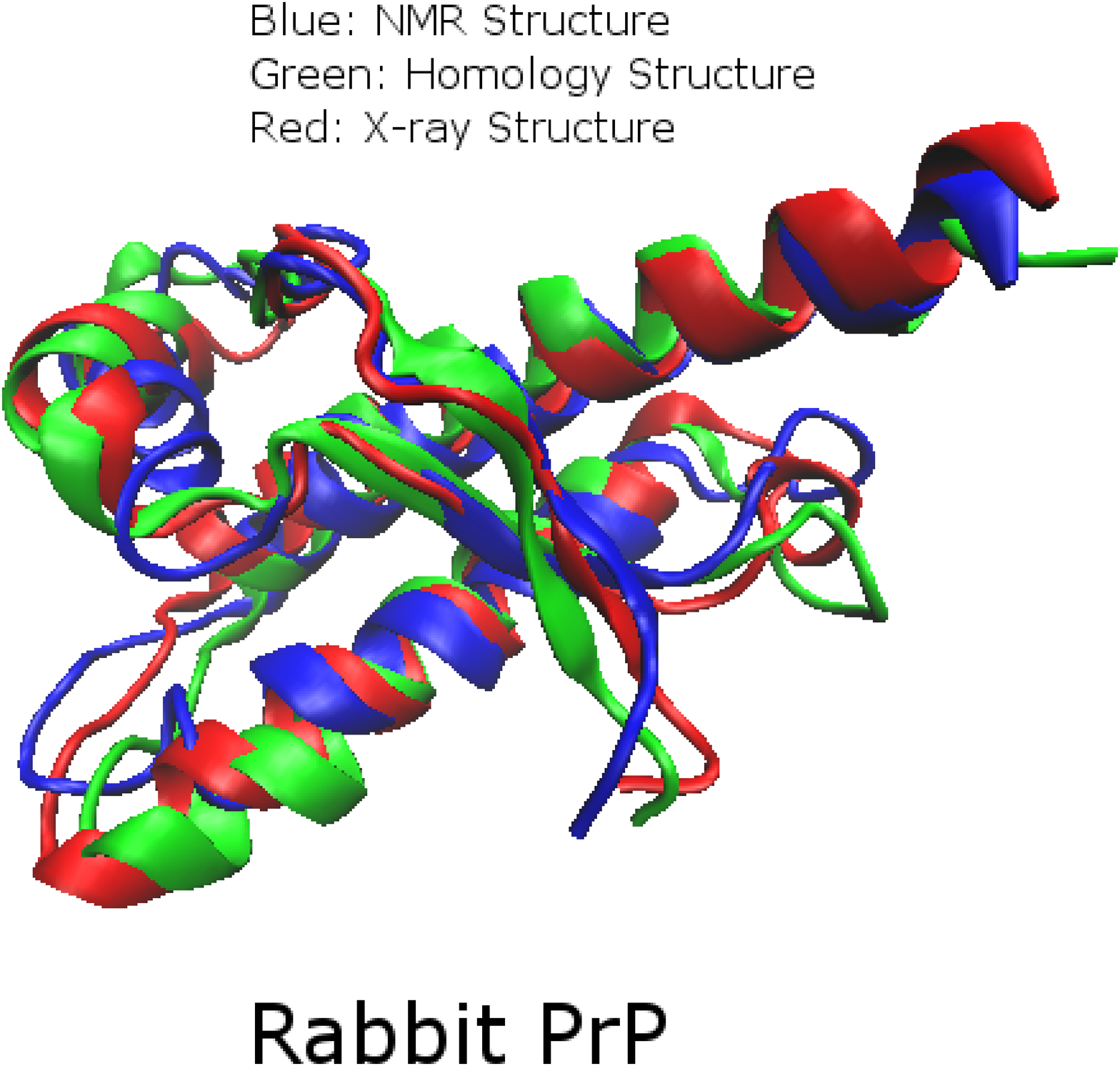} 
}
\caption{Rabbit prion protein NMR, homology and X-ray structures (2FJ3.pdb, 6EPA.pdb, and 3O79.pdb).}
\label{RaPrP_NMR-Homology-XRAY_superposed}
\end{figure}

Lastly, we illuminate the figure (Fig. \ref{RaPrP_NMR-Homology-XRAY_superposed}) of rabbit prion protein, including the homology, NMR and X-ray structures (6EPA.pdb, 2FJ3.pdb, and 3O79.pdb respectively). We superpose the homology structure onto the NMR structure and find the RMSD value is 3.2031669 angstroms. Similarly, we superpose the X-ray structure onto the NMR structure and we get their RMSD value is 2.7918559 angstroms. This implies the homology structure 6EPA.pdb made in 2004 by Epa \cite{zhangev2006} is as effective as the X-ray structure 3O79.pdb released recently on date 2010-11-24 (last modified on 2011-02-02).\\

\centerline{$<\text{Fig. \ref{RaPrP_NMR-Homology-XRAY_superposed}}>$}

\section{Conclusion}
To really reveal the secrets of prion diseases is very hard. Prion proteins have two regions: unstructured region and structured region. Rabbits, horses, and dogs were reported having immunity to prion diseases. Fortunately, by the end of 2010 all the NMR molecular structures of rabbit, horse, and dog prion proteins had been released into PDB bank already; for rabbit prion protein, its X-ray structure was also released into PDB bank in the end of 2010. Prion diseases are `structural conformational' diseases. This paper timely presents a clue to reveal some secrets in the view of the dynamics of prion molecular structures. MD experiences of the author nearly in the passing 10 years show to us a common conclusion: under low pH environment at many levels of temperatures with different starting MD velocities, the rabbit prion protein always unfolds $\alpha$-helical structures into $\beta$-sheet structures. Prion diseases are just caused by the conversion from predominant $\alpha$-helices of PrP$^{\text{C}}$ into rich $\beta$-sheets of PrP$^{\text{Sc}}$. Hence, we should furthermore study rabbits, horses and dogs, compared with humans and mice in order to reveal some secrets of prion diseases; for us, it is a long shot but certainly worth pursuing.\\

\noindent {\bf Acknowledgments:} This research is supported by a Victorian Life Sciences Computation Initiative (http://www.vlsci.org.au) grant number VR0063 on its Peak Computing Facility at the University of Melbourne, an initiative of the Victorian Government. The author appreciates kind invitations from the INTECH Open Access Publisher to write this book chapter.

\newpage
\begin{figure*}[h!]
\centerline{
\includegraphics[width=5.6in]{Fig1a.eps}
}
\end{figure*}

\newpage

\newpage

\newpage

\newpage

\newpage

\newpage

\newpage

\newpage
\begin{figure*}[h!]
\centerline{
\includegraphics[width=2.0in]{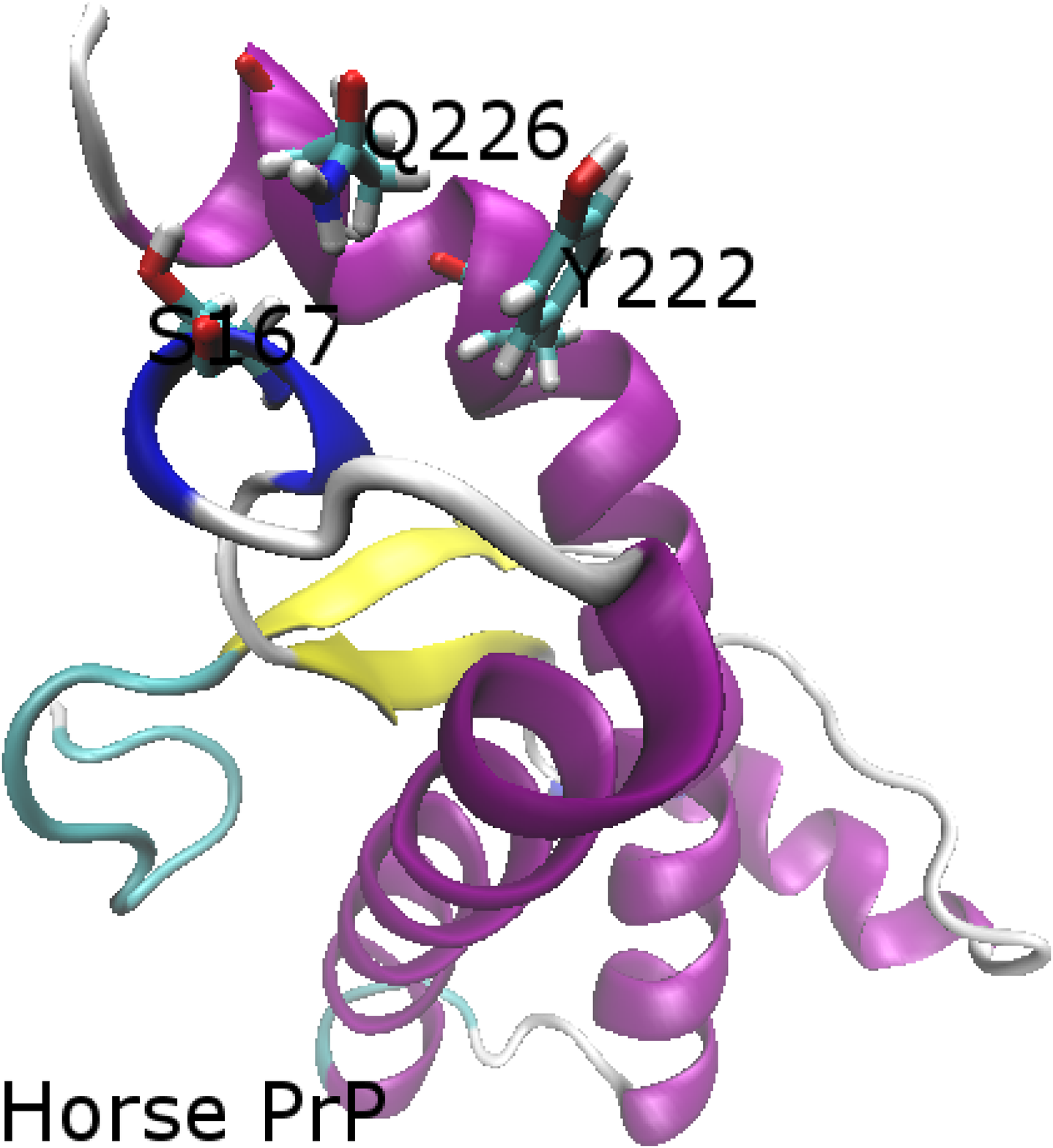}
\includegraphics[width=2.0in]{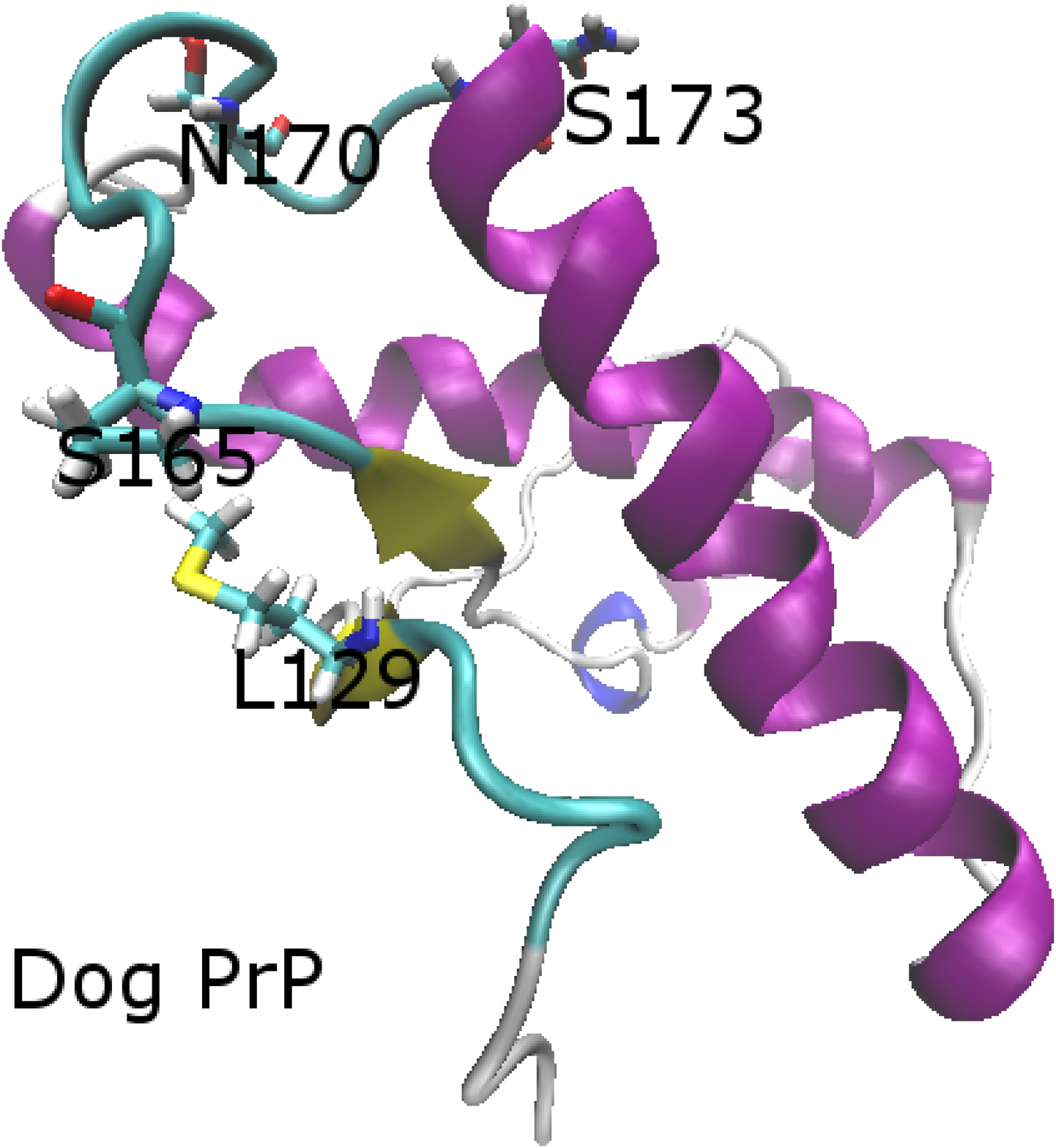}
\includegraphics[width=2.0in]{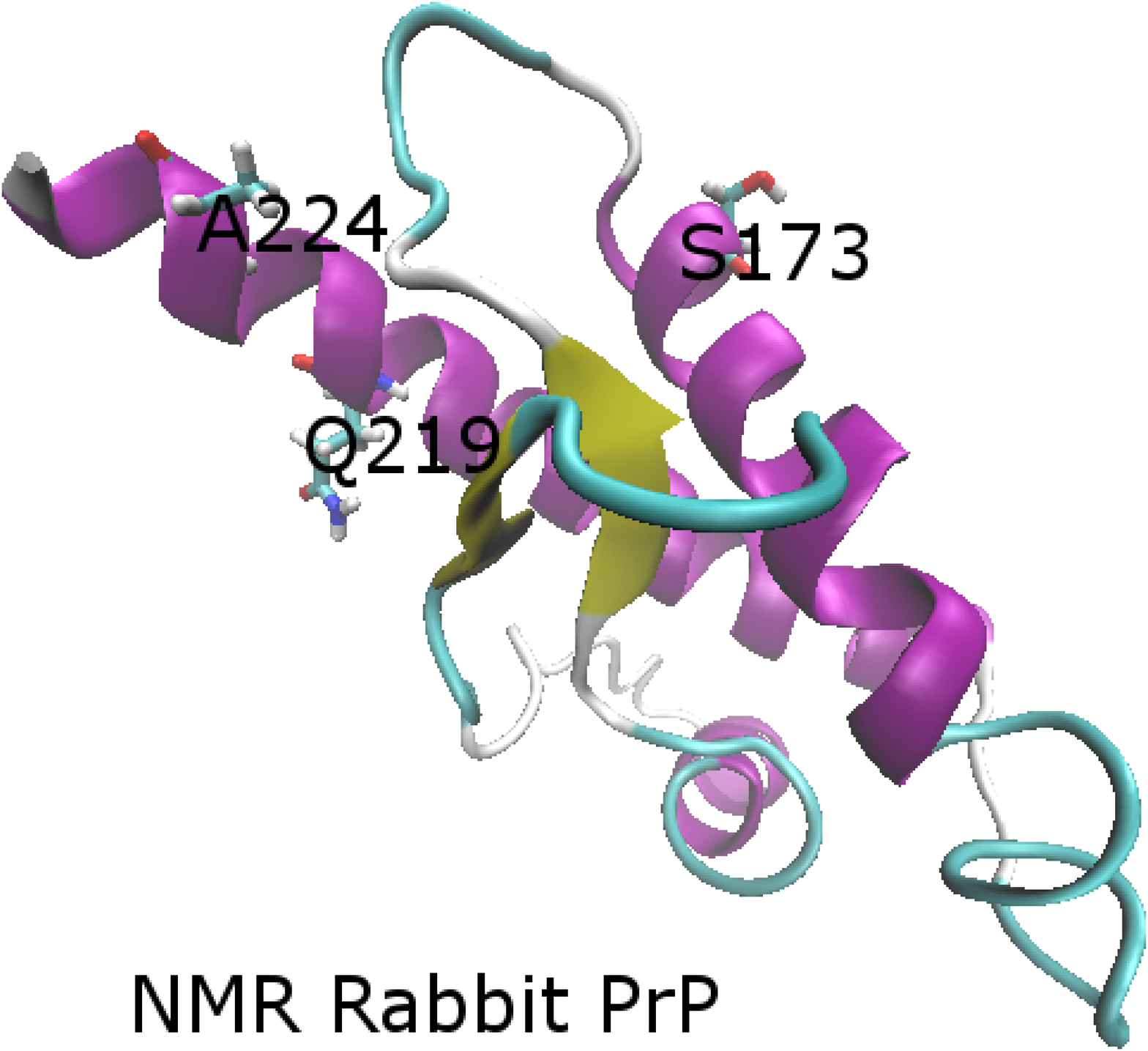}
}
\end{figure*}

\newpage

\end{document}